\documentclass[twocolumn,aps,pra,showpacs,superscriptaddress,tightenlines]{revtex4-1}
\usepackage{amsmath}
\usepackage{amsfonts}
\usepackage{graphicx}
\usepackage{epsfig}
\usepackage{color}
\usepackage[colorlinks,citecolor=blue]{hyperref}

\begin{document}
	
\title{Loss-induced quantum nonreciprocity}

\author{Baijun Li}
\affiliation{Institute of Quantum Precision Measurement, State Key Laboratory of Radio Frequency Heterogeneous Integration, College of Physics and Optoelectronic Engineering, Shenzhen University, Shenzhen 518060, China}

\affiliation{Quantum Science Center of Guangdong-Hongkong-Macao Greater Bay Area (Guangdong), Shenzhen 518045, China}

\author{Yunlan Zuo}
\affiliation{Key Laboratory of Low-Dimensional Quantum Structures
and Quantum Control of Ministry of Education, Department of
Physics and Synergetic Innovation Center for Quantum Effects and
Applications, Hunan Normal University, Changsha 410081, China}

\author{Le-Man Kuang}
\affiliation{Key Laboratory of Low-Dimensional Quantum Structures
and Quantum Control of Ministry of Education, Department of
Physics and Synergetic Innovation Center for Quantum Effects and
Applications, Hunan Normal University, Changsha 410081, China}

\author{Hui Jing}\email{jinghui73@foxmail.com}
\affiliation{Key Laboratory of Low-Dimensional Quantum Structures
and Quantum Control of Ministry of Education, Department of
Physics and Synergetic Innovation Center for Quantum Effects and
Applications, Hunan Normal University, Changsha 410081, China}

\author{Chaohong Lee}\email{chleecn@szu.edu.cn}
\affiliation{Institute of Quantum Precision Measurement, State Key Laboratory of Radio Frequency Heterogeneous Integration, College of Physics and Optoelectronic Engineering, Shenzhen University, Shenzhen 518060, China}

\affiliation{Quantum Science Center of Guangdong-Hongkong-Macao Greater Bay Area (Guangdong), Shenzhen 518045, China}
	
\begin{abstract}
Attribute to their robustness against loss and external noise, nonreciprocal photonic devices hold great promise for applications in quantum information processing.
Recent advancements have demonstrated that nonreciprocal optical transmission in linear systems can be achieved through the strategic introduction of loss.
However, a crucial question remains unanswered: can loss be harnessed as a resource for generating nonreciprocal quantum correlations?
Here, we take a counterintuitive stance by engineering loss to generate a novel form of nonreciprocal quantum correlations, termed \textit{nonreciprocal photon blockade}.
We examine a dissipative three-cavity system comprising two nonlinear cavities and a linear cavity.
The interplay of loss and nonlinearity leads to a robust nonreciprocal  single- and two-photon blockade, facilitated by destructive quantum interference.
Furthermore, we demonstrate the tunability of this nonreciprocal photon blockade by manipulating the relative phase between the two nonlinear cavities.
Remarkably, this allows for the reversal of the direction of nonreciprocity.
Our study not only sheds new light on the concept of loss-engineered quantum nonreciprocity but also opens up a unique pathway for the design of quantum nonreciprocal photonic devices.
\end{abstract}
		
\maketitle

\section{Introduction} \label{Int}
Nonreciprocal photonic devices play a vital role in quantum information processing due to their exceptional capability to effectively segregate backscattering noises from signals~\cite{Jalas13What}.
Beyond conventional magneto-optical nonreciprocal devices~\cite{Dotsch05Applications}, optical nonreciprocity can be achieved through magnet-free approaches, including time modulation~\cite{Sounas17Non}, non-Hermitian structure~\cite{Bender13Observation,Peng14Parity,Chang14Parity}, nonlinearity~\cite{Shi15Limitations,Shen16Experimental,Scheucher16Quantum,Tang22Quantum,Cotrufo24Passive}, artificial gauge field~\cite{Xu15Optical,Xu20Nonreciprocity,Chen21Synthetic}, and moving media~\cite{Maayani18Flying,Xu20Tunable,Wang22Acoustic}.
A recent and notable development in this field is the emergence of nonreciprocal photon blockade (PB)~\cite{Imamoglu97Strongly,Birnbaum05Photon,Huang18Nonreciprocal,Graf22Nonreciprocity,Yang23Non}, a novel form of directional quantum correlations, serving as a unique counterpart to classical nonreciprocity.
This phenomenon has evolved into a powerful tool for unidirectionally creating and manipulating few photons, enabling promising applications in single-photon isolators~\cite{Shen21Single,Dong21All}, circulator~\cite{Scheucher16Quantum}, and routers~\cite{Shomroni14All}.
The predictions of nonreciprocal PB span various systems, including spinning resoantors~\cite{Huang18Nonreciprocal,Li19Nonreciprocal,Wang19Nonreciprocal,Zhang23Nonreciprocal,Jing21Nonreciprocal,Bo23Spinning,Liu23Simultaneous}, asymmetric cavities~\cite{Xia21Giant,Gao23}, and chiral devices~\cite{Gu22Generation,Gao23Hybrid,Xie22Nonreciprocal,Liu23Parametric,Shen23Tunable}.
In experimental settings, nonreciprocal PB has been observed in a microdisk coupled with a scatterer~\cite{Graf22Nonreciprocity} and in a Fabry-P\'{e}rot cavity strongly coupled to atoms~\cite{Yang23Non}.
Moreover, the concept of quantum nonreciprocity extends widely across various fields, encompassing phenomena such as nonreciprocal phonon (magnon) blockade or lasing~\cite{Yao22Nonreciprocal,Yuan23Optical,Wang22Dissipation,Jiang18Nonreciprocal,Xu21Nonreciprocal,Huang22Parametric}, nonreciprocal solitons~\cite{Li21Nonreciprocal} or chaos~\cite{Zhang21Nonreciprocal}, nonreciprocal single-photon band structures~\cite{Tang22Nonreciprocal}, and one-way quantum ground-state cooling~\cite{Xu19Nonreciprocal,Lai20Nonreciprocal} or directional entanglement~\cite{Jiao20Nonreciprocal,Jiao22Nonreciprocal}.

Loss is an inherent aspect of realistic systems and often poses challenges in quantum information processing, traditionally considered a hindrance.
It is imperative to address these challenges and develop innovative approaches to protect quantum effects~\cite{Leghtas15Confining,Zhou22Environmentally,Asher23Non} against losses. In particular,
loss, typically perceived as detrimental, has been found to play a constructive role in recent experiments on loss-induced lasing revival~\cite{Peng14Loss}, and loss-assisted metasurface~\cite{Dong20Loss}.
Also, loss has been effectively employed to induce PB revival~\cite{Zuo22Loss}, transparency~\cite{Guo09Observation,Zhang18Loss,Zhang24Nonreciprocal}, non-Hermitian skin effects~\cite{Huang23Non}, and topological excitations~\cite{Pereira23Non}. Very recently, an elegant strategy has been devised to achieve nonreciprocal transmission with the assistance of loss~\cite{Huang21Loss,Huang23Loss} and a new field of loss-induced nonreciprocity is emerging.
However, current investigations predominantly concentrate on nonreciprocal optical transmission, while exploration of the quantum correlations of photons remains absent.
It remains unknown whether loss can serve as a resource for achieving nonreciprocal quantum correlations.

In this article, we introduce a counterintuitive mechanism to achieve nonreciprocal PB, a novel type of nonreciprocal quantum correlations, by leveraging the influence of loss.
Our study focuses on a dissipative three-cavity configuration consisting of two nonlinear cavities and a linear cavity.
Through the intricate interplay between Kerr nonlinearity and loss engineering, we discover that single-photon
blockade (1PB) characterized by a high transmission coefficient and two-photon blockade (2PB), can be selectively achieved by driving the system from one direction while failing from the other.
Furthermore, we demonstrate the ability to control the direction of nonreciprocity in both classical and quantum regimes by tuning the relative phase between the two nonlinear cavities.
This research marks an initial stride in the field of loss-engineered quantum nonreciprocity, not only offering advantages for chiral photonic communication and backaction-immune quantum sensing, but also advancing potential generalization to various quantum nonreciprocal effects in the future, such as, nonreciprocal macroscopic quantum superposition~\cite{Li23Optomechanical,Li23Nonreciprocal}, quantum squeezing~\cite{Guo23Nonreciprocal}, and another types of quantum correlations~\cite{Lu21Nonreciprocity}.

\begin{figure*}[t!]
  \includegraphics[width=\linewidth]{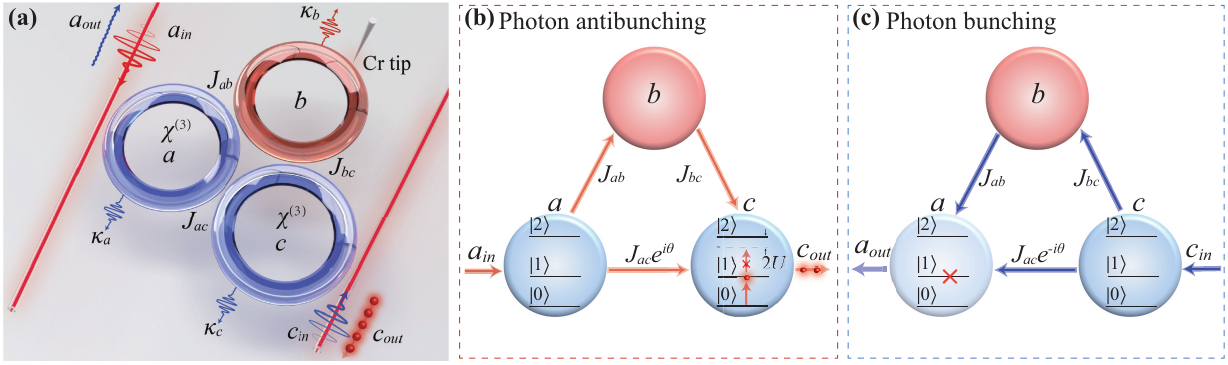}
  \centering
  \caption{{Nonreciprocal photon blockade (PB) induced by the loss of a auxiliary cavity.}
  {(a)} The system includes two nonlinear cavities $a,\ c$ with strong Kerr nonlinearity $\chi^{(3)}$ and a linear cavity $b$ with a engineered loss $\kappa_b$.
  The two nonlinear cavities are directly coupled with each other with the strength $J_{ac}$ and the phase factor $\theta$, and are coupled to a linear cavity with the strengths $J_{ab}$ and $J_{bc}$, respectively.
  $\kappa_a$, $\kappa_c$, and $\kappa_b$ are the loss strength of the corresponding cavities, respectively.
  (b, c) Physical mechanism illustration for nonreciprocal PB via the two-channel interference induced by the loss and the nonlinearity-induced anharmonicity of energy spectrum.
  For forward transmission ($a\to c$), photon is output from the cavity $c$ one by one because of the nonlinearity-induced anharmonicity of the eigenenergy spectrum.
  For backward transmission ($c\to a$), due to the destructive interference of photons in the two channels (i.e. the direct channel $c\to a$ and the indirect channel $c\to b \to a$), resulting in the significant reduction of one-photon population in cavity $a$, thus the output light exhibits bunching effect.}
  \label{Fig1}
\end{figure*}
%

\section{PHYSICAL SYSTEM AND STEADY-STATE BEHAVIOR}\label{sec2}
We consider a dissipative three-cavity system of two nonlinear cavities $a$ and $c$, and a linear resonator $b$, as shown in Figure~\ref{Fig1}(a).
In units of $\hbar=1$, the system can be described by the Hamiltonian
\begin{align}
\hat{H}=\ &\omega_a\hat{a}^\dag\hat{a}+\omega_c\hat{c}^\dag\hat{c}+\omega_b\hat{b}^\dag\hat{b}
+U_a\hat{a}^\dag\hat{a}^\dag\hat{a}\hat{a}+U_c\hat{c}^\dag\hat{c}^\dag\hat{c}\hat{c} \nonumber\\&
+(J_{ac}e^{i\theta}\hat{a}\hat{c}^\dagger+J_{ab}\hat{a}\hat{b}^\dagger+J_{bc}\hat{c}\hat{b}^\dagger+H.c.),
\label{Eq:1}
\end{align}
where $\hat{a}$, $\hat{c}$ and $\hat{b}$ are the annihilation operators for photons in the three resonators with resonance frequencies $\omega_a$, $\omega_c$ and $\omega_b$, respectively.
$J_{ac}$ is the coupling strength between cavities $a$ and $c$, and it contains a nonzero phase factor $\theta$.
In addition to considering directly complex coupling interactions, the coupling with phase factor can also be realized indirectly by using nonlinear and temporal modulation~\cite{Huang23Loss}.
$J_{ab}$ ($J_{bc}$) denotes the coupling strength between the nonlinear cavity $a$ ($c$) and the linear cavity $b$.
For whispering-gallery-mode resonator, the evanescent coupling of double resonators can be changed by adjusting the distance between them~\cite{Zhang18A}. $U_{a(c)}=\hbar\omega_{a(c)}^2c_0n_2/(n_0V_{eff})$ is the Kerr nonlinear parameter with the nonlinear (linear) refractive index $n_2\ (n_0)$, the speed of light in vacuum $c_0$, and the effective mode volume of resonator $V_{eff}$.

In order to detect the nonreciprocal characters of the system, a weak driving laser with frequency $\omega_d$ and amplitude $\Omega$ is input into the cavity $a$ or cavity $c$.
Thus the driving term is described by
\begin{align}
\hat{H}_d=\Omega(\hat{d}^\dagger e^{-i\omega_pt}+\hat{d}e^{i\omega_pt}),
\label{Eq:2}
\end{align}
where $\hat{d}=\hat{a}$ or $\hat{c}$.
In a rotating frame with respect to the driving frequency $\omega_p$, the total Hamiltonian becomes
\begin{align}
\hat{H_t}=\ &\Delta_a\hat{a}^\dag\hat{a}+\Delta_c\hat{c}^\dag\hat{c}+\Delta_b\hat{b}^\dag\hat{b}
+U_a\hat{a}^\dag\hat{a}^\dag\hat{a}\hat{a}
+U_c\hat{c}^\dag\hat{c}^\dag\hat{c}\hat{c} \nonumber\\&
+(J_{ac}e^{i\theta}\hat{a}\hat{c}^\dagger+J_{ab}\hat{a}\hat{b}^\dagger
+J_{bc}\hat{c}\hat{b}^\dagger+H.c.) \nonumber\\&
+\Omega(\hat{d}^\dagger +\hat{d}),
\label{Eq:3}
\end{align}
where $\Delta_{a(c,b)}=\omega_{a(c,b)}-\omega_p$ is the frequency detuning between the cavity mode and the driving field.
Thus the time-evolution dynamics obeys the master equation~\cite{Walls94Quantum,Johansson13Qutip,Gardiner00Quantum},
\begin{align}
\frac{\partial \hat{\rho}}{\partial t} =\ &-i[\hat{H}_t, \hat{\rho}] + \frac{\kappa_a}{2} \mathcal{L}[\hat{a}]\hat{\rho} + \frac{\kappa_c}{2} \mathcal{L}[\hat{c}]\hat{\rho}
+ \frac{\kappa_b}{2} \mathcal{L}[\hat{b}]\hat{\rho},
\label{Eq:9}
\end{align}
where $\hat{\rho}$ is the density operator and  $\mathcal{L}[\hat{o}]=2\hat{o}\hat{\rho}\hat{o}^\dagger-(\hat{o}^\dagger\hat{o}\hat{\rho}+\hat{\rho}\hat{o}^\dagger\hat{o})$ describes a Lindbland term for operator $\hat{o}$.
$\kappa_a$, $\kappa_c$ and $\kappa_b$, are loss strengths for the three cavities,  respectively.
The loss strength $\kappa_b$ can be engineered by placing a chromium-coated silica-nanofiber tip in the vicinity of the microcavity $b$, featuring strong absorption in the 1550 nm band~\cite{Peng14Loss}.
The strength of absorption can be enlarged by increasing the overlap volume between the cavity mode and the nanotip, resulting in a widening of the linewidth without a significant change in resonance frequency~\cite{Peng14Loss}.
In the above master equation, the thermal effects have been neglected due to the high resonant frequencies for the three optical cavities.

To understand the flow of photons, one can calculate the transmission coefficients
\begin{align}
T_{a\to c} = & \frac{\langle\hat{c}_{out}^\dagger\hat{c}_{out}\rangle}{P_{in}^a}=\frac{\kappa_a\kappa_c}{\Omega^2}\langle\hat{c}^\dagger\hat{c}\rangle, \nonumber\\\ T_{c\to a} = & \frac{\langle\hat{a}_{out}^\dagger\hat{a}_{out}\rangle}{P_{in}^c}=\frac{\kappa_a\kappa_c}{\Omega^2}\langle\hat{a}^\dagger\hat{a}\rangle,
\label{Eq:4}
\end{align}
for the photons transmitted from cavity $a\ (c)$ to cavity $c\ (a)$.
Here $\langle \cdot \rangle$ denotes quantum expectation with respect to a steady state and we have utilized the the input-output relation~\cite{Gardiner85Input}.
When the system is driven from the left side (cavity $a$), we have $P_{in}^a=\Omega^2/\kappa_a$ for the input field and $\hat{c}_{out}=\sqrt{\kappa_c}\hat{c}$ for the output field.
While the system is driven from the right side (cavity $c$), we have $P_{in}^c=\Omega^2/\kappa_c$ and $\hat{a}_{out}=\sqrt{\kappa_a}\hat{a}$.
If the perfect transmission from cavity $a$ to cavity $c$ takes place, we have $T_{a\to c}=1$, and vice versa.

To characterize the quantum correlations among the transmitted photons, one can calculate the equal-time second-order and third-order correlation functions in the steady state~\cite{Walls94Quantum}
\begin{align}
g_{a\to c}^{(n)}(0) &= \frac{\langle\hat{c}_{out}^{\dagger n} \hat{c}_{out}^{n}\rangle}{\langle\hat{c}_{out}^\dagger\hat{c}_{out}\rangle^n}
=\frac{\langle\hat{c}^{\dagger n}\hat{c}^{n}\rangle}{\langle\hat{c}^\dagger\hat{c}\rangle^n}, \nonumber \\
g_{c\to a}^{(n)}(0) &= \frac{\langle\hat{a}_{out}^{\dagger n} \hat{a}_{out}^{n}\rangle}{\langle\hat{a}_{out}^\dagger\hat{a}_{out}\rangle^n}
=\frac{\langle\hat{a}^{\dagger n}\hat{a}^{n}\rangle}{\langle\hat{a}^\dagger\hat{a}\rangle^n},
\label{Eq:6}
\end{align}
for the photons output from cavity $c$ and cavity $a$, respectively, and $n=2,\ 3$.
Given a steady state, $g^{(2)}_{a\to c}(0)\ [g^{(2)}_{c\to a}(0)]>1$ and $g^{(2)}_{a,c}(0)\ [g^{(2)}_{c\to a}(0)]<1$ correspond to super-Poissonian statistics (bunching) and sub-Poissonian statistics (antibunching), respectively.
In the case of perfect 1PB, we have $g^{(2)}_{a\to c}(0)\ [g^{(2)}_{c\to a}(0)] \to 0$~\cite{Walls94Quantum}.
For the 2PB takes place, we have $g^{(2)}_{a\to c}(0)\ [g^{(2)}_{c\to a}(0)]>1$ and $g^{(3)}_{a\to c}(0)\ [g^{(3)}_{c\to a}(0)]<1$.

\begin{figure*}
\includegraphics[width=\linewidth]{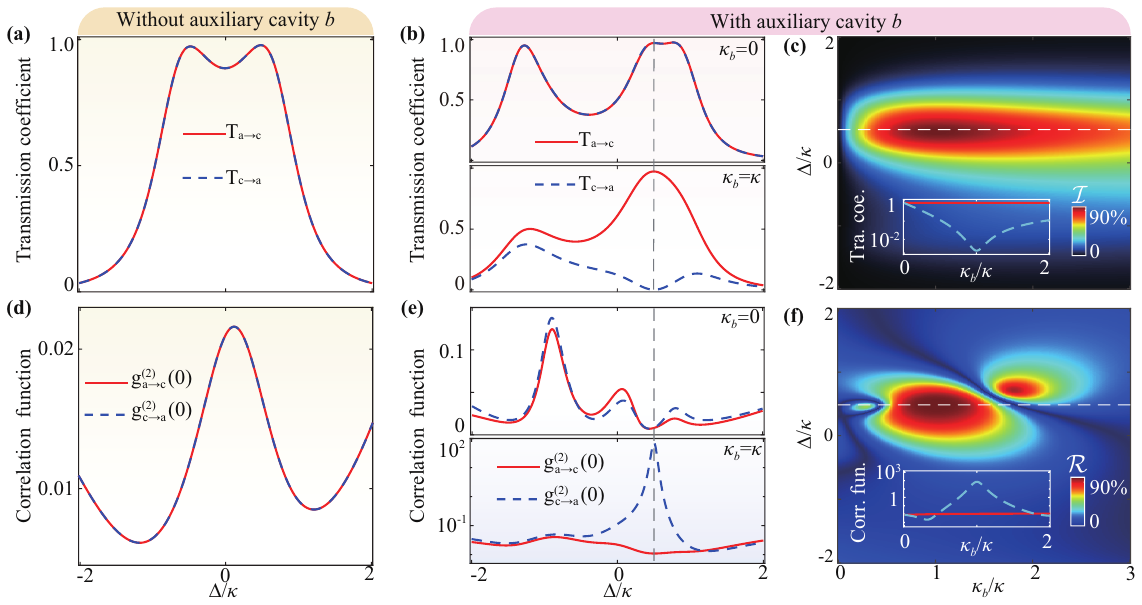}
  \caption{Classical nonreciprocal transmission and quantum nonreciprocal 1PB induced by the loss of cavity $b$.
  {(a)} {Transmission coefficients $T_{a\to c}$ (red solid curve) and $T_{c\to a}$ (blue dashed curve) as a function of the detuning $\Delta/\kappa$ in a two-cavity nonlinear system without the auxiliary cavity $b$.}
  {(b)} {$T_{a\to c}$ and $T_{c\to a}$ versus the detuning $\Delta/\kappa$ for different  loss strengths of the auxiliary cavity $b$ ($\kappa_b/\kappa=0,1$) in a three-cavity nonlinear system.}
  {(c)} {Classical isolation $\mathcal{I}=|T_{a\to c}-T_{c\to a}|$ as a function of the detuning $\Delta/\kappa$ and the loss $\kappa_b/\kappa$.
  The inset shows the $T_{a\to c}$ and $T_{c\to a}$ versus the loss $\kappa_b/\kappa$ for a fixed detuning $\Delta=0.5\kappa$}.
  {(d)} {Equal-time second-order correlation functions $g^{(2)}_{a\to c}(0)$ and $g^{(2)}_{c\to a}(0)$ versus the detuning $\Delta/\kappa$ in the two-cavity system.
  {(e)} $g^{(2)}_{a\to c}(0)$ and $g^{(2)}_{c\to a}(0)$ versus the detuning $\Delta/\kappa$ for different loss strengths $\kappa_b$ ($\kappa_b/\kappa=0,1$) in the three-cavity system.}
  {(f)} {Quantum nonreciprocal ratio $\mathcal{R}$ as a function of the $\Delta/\kappa$ and the loss $\kappa_b/\kappa$. The inset shows $g^{(2)}_{a\to c}(0)$ and $g^{(2)}_{c\to a}(0)$ versus the loss $\kappa_b/\kappa$  for a fixed detuning $\Delta=0.5\kappa$}.
  {The other parameters are chosen as $\kappa_a=\kappa_c=\kappa$, $J_{ab}=J_{bc}=J_{ac}=\sqrt{2}\kappa/2$, $\Omega=0.1\kappa$, $U_{a}=U_{c}=5\kappa$, and $\theta=-\pi/4$.}}
  \label{Fig2}
\end{figure*}

For a nonlinear system, it is very hard to solve the equations~(\ref{Eq:4},\ref{Eq:6}) analytically.
Thus, we numerically simulate the dynamics by using the qutip~\cite{Johansson13Qutip}.
For a weak pump field ($\Omega\ll \{\kappa_a, \kappa_c\}$), the population in one-photon states is much larger than the one in two-photon states, which are almost unaffected by the nonlinearity of the system.
Therefore, the Kerr term in Hamiltonian~(\ref{Eq:1}) can be ignored in the transmission behavior, and one can analytically calculate the transmission coefficient of the system in the frequency domain~\cite{Xu15Optical,Xu20Nonreciprocity}.
According to the method given in~\cite{Gardiner00Quantum}, one can derive the quantum Langevin equations for the annihilation operators
\begin{align}
\frac{d}{dt}\hat{a}=&-\left(i\omega_a+\frac{\kappa_a}{2} \right)\hat{a}-iJ_{ac}e^{-i\theta}\hat{c}-iJ_{ab}\hat{b}+\sqrt{\kappa_a}\hat{a}_{in},
\nonumber \\ \frac{d}{dt}\hat{c}=& -\left(i\omega_c+\frac{\kappa_c}{2} \right)\hat{c}-iJ_{ac}e^{i\theta}\hat{a}-iJ_{bc}\hat{b}+\sqrt{\kappa_c}\hat{c}_{in},
\nonumber \\ \frac{d}{dt}\hat{b}=& -\left(i\omega_b+\frac{\kappa_b}{2} \right)\hat{b}-iJ_{ab}\hat{b}-iJ_{bc}\hat{c}+\sqrt{\kappa_b}\hat{b}_{in},
\label{Eq:11}
\end{align}
where $\hat{a}_{in}, \hat{c}_{in}$, and $\hat{b}_{in}$ are the input operators within environments noises.

By introducing the vectors of the operators,
\begin{align}
\hat{u}^{T}=&\ (\hat{a}, \hat{c}, \hat{b}), v_{in}^{T}=(\hat{a}_{in}, \hat{c}_{in}, \hat{b}_{in}),  \Gamma = \mathrm{diag}(\kappa_a, \kappa_c, \kappa_b),
\label{Eq:12}
\end{align}
we can rewritten the Langevin equations as
\begin{align}
\frac{d}{dt}\hat{u}=-iMu+\sqrt{\Gamma}v_{in},
\label{Eq:13}
\end{align}
with
\begin{align}
M=\begin{pmatrix}
      \omega_a-i\frac{\kappa_a}{2} & J_{ac}e^{-i\theta} & J_{ab}\\
			J_{ac}e^{i\theta} & \omega_c-i\frac{\kappa_c}{2} & J_{bc}\\
			J_{ab} & J_{bc} & \omega_b-i\frac{\kappa_b}{2}
       \end{pmatrix}.
\label{Eq:141}
\end{align}
By using the Fourier transformation, the solution of equation~(\ref{Eq:13}) can be given as
\begin{align}
u(\omega)=-i(M-\omega I)^{-1}\sqrt{\Gamma}v_{in}(\omega),
\label{Eq:152}
\end{align}
where $I$ is the unit matrix.
Using the input-output relation in the frequency domain, $u_{out}(\omega)=-u_{in}(\omega)+\sqrt{\Gamma}u(\omega)$, the scattering coefficient matrix of the system reads
\begin{align}
S(\omega)=\frac{u_{out}}{u_{in}}=-i\sqrt{\Gamma}(M-\omega I)^{-1}\Gamma - I.
\label{Eq:14}
\end{align}
Setting $\omega_a=\omega_c=\omega_b=\omega_0$ and $J_{ab}=J_{bc}=J$, the transmission coefficients between cavity $a$ and cavity $c$ are given as
\begin{align}
T_{a\to c}(\omega)=&\left|S_{21}(\omega) \right|^2 \nonumber \\=& \left|\frac{-i\sqrt{\kappa_a\kappa_c} \left[J^2+e^{i\theta}J_{ac}\left( \delta+\frac{i\kappa_b}{2}\right)  \right]}{D(\omega)}   \right|^2,
\label{Eq:15}
\end{align}
and
\begin{align}
T_{c\to a}(\omega)=&\left|S_{12}(\omega) \right|^2 \nonumber \\=&\left|\frac{-i\sqrt{\kappa_a\kappa_c}\left[J^2+e^{-i\theta}J_{ac}\left( \delta+\frac{i\kappa_b}{2}\right)  \right]}{D(\omega)}   \right|^2,
\label{Eq:151}
\end{align}
where $\delta=\omega-\omega_0$ and
\begin{align}
D(\omega)=\ &2\mathrm{cos}(\theta)J^2J_{ac}+J_{ac}^2\left(\delta+\frac{i\kappa_b}{2} \right)\nonumber \\&+J^2\left(\delta+\frac{i\kappa_c}{2} \right)-\left(\delta+\frac{i\kappa_a}{2} \right)\nonumber \\
&\times
\left[-J^2+\left(\delta+\frac{i\kappa_b}{2} \right)\left(\delta+\frac{i\kappa_c}{2} \right)  \right].
\label{Eq:16}
\end{align}

\begin{figure}[t!]
  \includegraphics[width=0.9\linewidth]{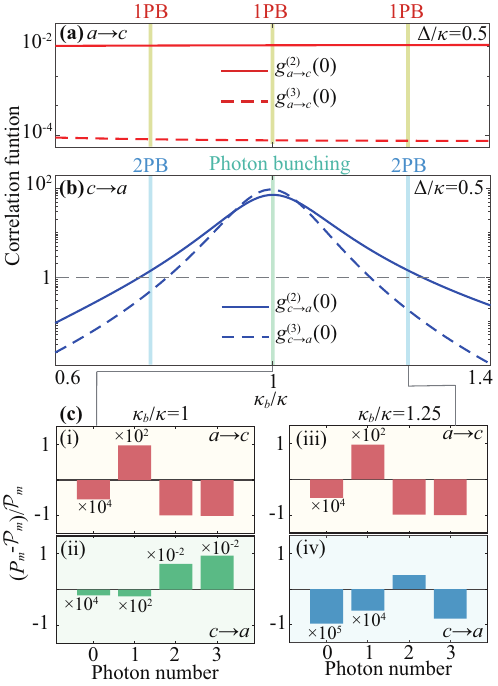}
  \centering
  \caption{{Nonreciprocal 2PB engineered by the loss of cavity $b$.}
  {(a, b)} {Equal-time second-order [$g^{(2)}_{a\to c}(0)$, $g^{(2)}_{c\to a}(0)$] and third-order [$g^{(3)}_{a\to c}(0)$, $g^{(3)}_{c\to a}(0)$] correlation functions as functions of the loss $\kappa_b/\kappa$ for a fixed detuning $\Delta=0.5\kappa$.
   {(c)} {The deviations of the photon distribution $P_m$ to the standard Poisson distribution $\mathcal{P}_m$ versus the photon number $m$ at a fixed detuning $\Delta=0.5\kappa$ and different loss strengths $\kappa_b/\kappa=1,\ 1.25$.
  The other parameters are the same ones in Figure~\ref{Fig2}.}}}
  \label{Fig3}
\end{figure}

According to the transmission coefficients~(\ref{Eq:15}), one can find the optimal parameter condition for the ideal nonreciprocal transmission.
It is also an essential prerequisite for the preparation of nonreciprocal PB.
Under the condition of $\kappa_a=\kappa_c=\kappa$, the complete transmission from cavity $a$ to cavity $c$ (i.e. $T_{a\to c}=1$ and $T_{c\to a}=0$) takes place when
\begin{align}
\delta=\frac{\kappa}{2\mathrm{tan}(\theta)}, J_{ac}=-\frac{\kappa}{2\mathrm{sin}(\theta)},
 J= \pm J_{ac}, \kappa_b=\kappa,
\label{Eq:17}
\end{align}
and the complete transmission from cavity $c$ to cavity $a$ (i.e. $T_{a\to c}=0$ and $T_{c\to a}=1$) occurs when
\begin{align}
\delta=-\frac{\kappa}{2\mathrm{tan}(\theta)}, J_{ac}=\frac{\kappa}{2\mathrm{sin}(\theta)},
J= \pm J_{ac}, \kappa_b=\kappa.
\label{Eq:18}
\end{align}

\section{LOSS-INDUCED NONRECIPROCAL PHOTON BLOCKADE}\label{sec3}
We first analyze the light transmission.
To precisely obtain the transmission coefficients, we numerically solve the equations~(\ref{Eq:4}).
We assume parameter as $\Delta_a=\Delta_c=\Delta_b=\Delta$, $\kappa_a=\kappa_c=\kappa$, $U_{a}=U_{c}=5\kappa$, $\Omega=0.1\kappa$ and $\theta=-\pi/4$, and use optimal parameter condition $J_{ab}=J_{bc}=J_{ac}=\sqrt{2}\kappa/2$.
If the auxiliary cavity $b$ is absent, the system consists of two identical nonlinear resonators coupled to each other.
The transmission coefficients $T_{a\to c}$ (red solid curve) and $T_{c\to a}$ (blue dashed curve) versus the detuning $\Delta/\kappa$ are shown in Figure~\ref{Fig2}(a).
Obviously, nonreciprocity is absent in such a two-cavity system even with strong Kerr nonlinearity, i.e., $T_{a\to c}=T_{c\to a}$.
This is because that photons are occupied predominantly in one-photon state, which is nearly unaffected by the Kerr nonlinearity.

Surprisingly, by introducing a auxiliary lossy cavity $b$, the three-cavity system may exhibit huge nonreciprocity among the transmitted photons.
As shown in Figure~\ref{Fig2}(b), for a three-cavity system with $\kappa_b\approx\kappa$, photons can be almost completely transmitted from cavity $a$ to cavity $c$ (i.e. $T_{a\to c}\approx1$), but there is almost no photon transmission in the reverse direction (i.e. $T_{c\to a}\approx0$).
To show the loss-induced nonreciprocal transmission more clearly, we introduce the isolation $\mathcal{I}=|T_{a\to c}-T_{c\to a}|$ and show $\mathcal{I}$ as a function of $\Delta/\kappa$ and $\kappa_b/\kappa$ in the  in Figure~\ref{Fig2}(c).
The maximum isolation occurs at the points of $\Delta/\kappa\approx0.5$ and $\kappa_b/\kappa=1$, consistent with the results of the optimal conditions in equation~(\ref{Eq:17}).
Physically, this nonreciprocity stems from the destructive interference of the two transmission channels: $c\to a$ and $c\to b \to a$, leading to the sharp reduction of one-photon population in cavity $a$ and $T_{c\to a}\approx0$.

\begin{figure}[t!]
  \includegraphics[width=0.9\linewidth]{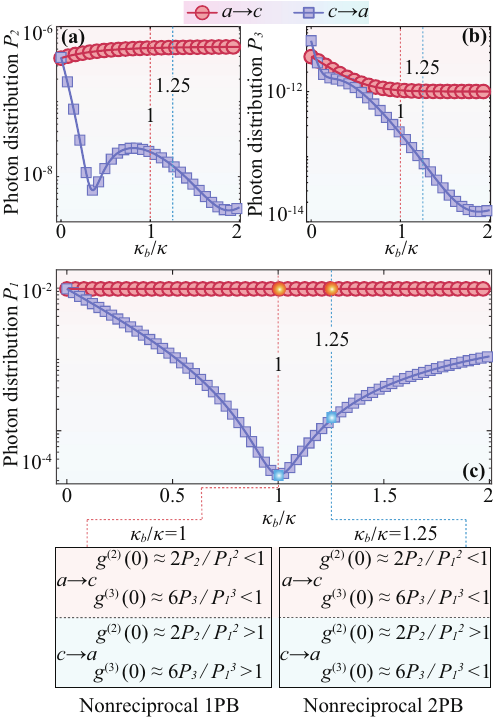}
  \centering
  \caption{{Inhomogeneous reduction of photon population in different photon states in the bidirectional photon transmission.}
  {(a-c)} {Photon distributions $P_m$ ($m=1,2,3$) versus the loss of auxiliary cavity $\kappa_b/\kappa$ for a fixed detuning $\Delta=0.5\kappa$ in the bidirectional transmission of photons.
  The other parameters are the same ones in Figure~\ref{Fig2}.}}
  \label{Fig4}
\end{figure}

By analysing the numerator of equation~(\ref{Eq:15}), one can find a straightforward condition for nonreciprocal transmission: $J_{ac}|\delta+i\kappa_b/2|=J^2$, $\phi-\theta=\pi+2n\pi$ or $\phi+\theta=\pi+2n\pi$, where $\phi=\mathrm{arg}(\delta+i\kappa_b/2)$ is the phase lag~\cite{Huang21Loss} induced by the loss and $n$ is the integer.
The phase lag $\phi$ is the dominant factor in the phase matching condition for the destructive two-channel interference, and it is primarily dependent of the ratio between loss and detuning of the cavity $b$.
Thus the loss engineering of the auxiliary cavity can provide an enticing avenue for manipulating the nonreciprocity.
The loss-induced nonreciprocal  optical transmission~\cite{Huang21Loss,Huang23Loss} has also been predicted in a linear multi-mode system.


Besides nonreciprocal optical transmission, nonreciprocal quantum correlation may appear as a result of combining loss engineering and Kerr nonlinearity.
We analyze the equal-time second-order and third-order correlation functions under the optimal parameters [Equation~(\ref{Eq:17})].
Figures~\ref{Fig2}(d-f) show the equal-time second-order correlation functions $g^{(2)}_{a\to c}(0)$ and $g^{(2)}_{c\to a}(0)$ in a two-cavity device and a three-cavity system.
If the loss engineering of cavity $b$ is absent, that is, cavity $b$ is absent or $\kappa_b=0$, as shown in Figure~\ref{Fig2}(d, e), the $g^{(2)}_{a\to c}(0)$ and $g^{(2)}_{c\to a}(0)$ are reciprocal regardless of the transmission directions.
At the same time, the strong 1PB [$\{g^{(2)}_{a\to c}(0), g^{(2)}_{c\to a}(0)\}\ll1$] may appear in both driving directions.
The physical mechanism of the 1PB is the anharmonic energy spectra of cavities ($a, c$) caused by the Kerr nonlinearity.
The resonance absorption of one photon will block the entry of subsequent photons, thus the second photon is excluded from the cavity.

In contrast, for the system with the loss engineering, $\kappa_b=\kappa$, the second-order correlation $g^{(2)}_{a\to c}(0)$ and $g^{(2)}_{c\to a}(0)$ can exhibit giant nonreciprocity, see Figure~\ref{Fig2}(e).
The 1PB can be generated [i.e., $g^{(2)}_{a\to c}(0)\ll1$], for the photons transmitted from cavity $a$ to $c$, whereas it is almost completely suppressed [i.e., $g^{(2)}_{c\to a}(0)\gg1$] in the reverse direction.
To see the loss-induced nonreciprocal quantum correlation more clearly, we introduce the quantum nonreciprocal  ratio $\mathcal{R}=|(g^{(2)}_{a\to c}(0)-g^{(2)}_{c\to a}(0))/(g^{(2)}_{a\to c}(0)+g^{(2)}_{c\to a}(0))|$.
The $\mathcal{R}=1\ (\mathcal{R}=0)$ corresponds to the ideal nonreciprocal 1PB (the disappearance of the quantum nonreciprocity).
The $\mathcal{R}$ versus the $\Delta/\kappa$ and the $\kappa_b/\kappa$ is shown in Figure~\ref{Fig2}(f).
We find that the locations where perfect nonreciprocal 1PB occurs are essentially the same as the ideal nonreciprocal transmission.
It implies that the strong 1PB is accompanied by a high transmission coefficient.
At the same time, the significant quantum nonreciprocity with difference in correlation functions up to 5 orders of magnitude for photons transmitted from opposite directions, see the inset graph in Figure~\ref{Fig2}(f).

By further refining the loss parameter, nonreciprocal 2PB can also be achieved.
In Figures~\ref{Fig3}(a, b), we plot the second-order $g^{(2)}_{a\to c}(0)$ and $g^{(2)}_{c\to a}(0)$, and third-order correlation functions $g^{(3)}_{a\to c}(0)$ and $g^{(3)}_{c\to a}(0)$ as functions of the loss $\kappa_b/\kappa$ at a fixed detuning $\Delta/\kappa=0.5$.
Fixing the loss around $\kappa_b/\kappa=1.25$, we find that 1PB [$g^{(2)}_{a\to c}(0)<1$] emerges by driving the system from the left side, while we have 2PB [$g^{(2)}_{c\to a}(0)>1$ and $g^{(3)}_{c\to a}(0)<1$] by driving  the system from the right side.
The nonreciprocal 1PB is further reproduced at $\kappa_b/\kappa=1$.
The nonreciprocity of 1PB and 2PB can also be confirmed by comparing the photon-number distribution $P_{a,c}(m)$ and the Poisson distribution $\mathcal{P}_{a,c}(m)$.
From Figures~\ref{Fig3}(c) (iv), we find that $P_{a}(2)$ is enhanced while $P_{a}(m>2)$ are suppressed by driving the system from right side.
As a sharp contrast, only $P_{c}(1)$ is enhanced in the opposite drive direction [see Figures~\ref{Fig3}(c) (iii)].
Similar phenomenon for nonreciprocal 1PB can be found in Figures~\ref{Fig3}(c) (i, ii).

\begin{figure}[t!]
  \includegraphics[width=0.9\linewidth]{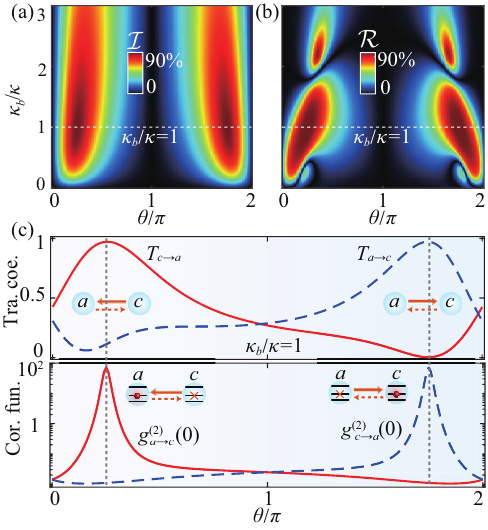}
  \caption{Reversing the direction of nonreciprocity by manipulating the relative phase $\theta$ between the two nonlinear cavities.
  {Transmission isolation $\mathcal{I}$ and quantum nonreciprocal ratio $\mathcal{R}$ as functions of the detuning $\Delta/\kappa$ and the loss $\kappa_b/\kappa$.
  {(c)} {Transmission coefficients $T_{a\to c}$ (red solid curve) and $T_{c\to a}$ (blue dashed curve), and second-order correlation functions $g^{(2)}_{a\to c}(0)$ (red solid curve) and $g^{(2)}_{c\to a}(0)$ (blue dashed curve) versus the preset phase $\theta/\pi$ for a fixed detuning $\Delta=0.5\kappa$ and loss $\kappa_b=\kappa$}.
  The other parameters are the same ones in Figure~\ref{Fig2}.}}
  \label{Fig5}
\end{figure}

The physical mechanism of nonreciprocal PB can be explained as follows.
Since the same anharmonicity of the energy spectrum exists in two nonlinear resonators ($a$ and $c$), two-channel quantum interference is responsible for observing strong nonreciprocal PB.
The loss engineering of the auxiliary cavity $b$ can lead to different phase matching condition for the destructive interference in the bidirectional photons transmission.
By driving the cavity $a$, see Figure~\ref{Fig1}(b), due to lack of the perfect phase matching condition for destructive quantum interference in the indicated cavities, the photons are transmitted to cavity $c$ completely.
And the 1PB [$g^{(2)}_{a\to c}(0)\ll1$] appears due to the anharmonicity of the energy spectrum in the cavity $c$.
In contrast, when the system is driven from the opposite direction, see Figure~\ref{Fig1}(c), the phase match condition is satisfied.
The direct photons transmission from cavity $c$ to $a$ will be reduced by the destructive quantum interference with the indirect paths, i.e., $c\to b \to a$, thus the output light can exhibit bunching effect.

In Figure~\ref{Fig4}, we further show the  photon distribution $P_m$ ($m=1,2,3$) versus the loss $\kappa_b/\kappa$ for a fixed detuning $\Delta=0.5\kappa$ in the bidirectional transmission of photons.
As shown in Figure~\ref{Fig4}(c), due to  the destructive quantum interference,  the sharp reduction of population is mainly reflected in the one-photon state ($P_1$) in cavity $a$. At the same time, the multi-photon state population ($P_2,\ P_3$) are not reduced equally, shown in Figure~\ref{Fig4}(a, b), thus the output light for the cavity $a$ exhibits bunching effect, i.e., $g^{(2)}_{c\to a}(0)\approx2P_2^a/(P_1^a)^2\gg1$.
Similarly, the nonreciprocal 2PB, $g^{(2)}_{c\to a}(0)\approx2P_2^a/(P_1^a)^2>1$, $g^{(3)}_{c\to a}(0)\approx6P_2^a/(P_1^a)^3<1$ and $g^{(2)}_{a\to c}(0)\approx2P_2^c/(P_1^c)^2<1$, also originates from different reductions of the photon  population in different photon states, modulated by loss engineering.

By changing the preset phase $\theta$, the direction of nonreciprocity in both classical and quantum regimes can also be reversed under the loss engineering.
In Figures~\ref{Fig5}(a,b), we show the classical isolation $\mathcal{I}$ and quantum nonreciprocal ratio $\mathcal{R}$ versus the detuning $\Delta/\kappa$ and the loss $\kappa_b/\kappa$.
We find that the maximum $\mathcal{I}$ and $\mathcal{R}$ corresponds to two phases $\theta$ for a optimal $\kappa_b$.
To understand this phenomenon, the transmission coefficients and second-order correlation functions for both directions are further plotted as functions of preset phase $\theta$ at a fixed $\Delta=0.5\kappa$ and $\kappa_b=\kappa$ in Figures~\ref{Fig5}(c).
We find $T_{a\to c}\approx0$ and $T_{c\to a}\approx1$ around phase $\theta=\pi/4$, and $T_{a\to c}\approx1$ and $T_{c\to a}\approx0$ with phase $\theta=3\pi/4$.
More importantly, we get $g^{(2)}_{a\to c}(0)\gg1$ and $g^{(2)}_{c\to a}(0)\ll1$ for phase $\theta=\pi/4$, and $g^{(2)}_{a\to c}(0)\ll1$ and $g^{(2)}_{c\to a}(0)\gg1$ with phase $\theta=3\pi/4$.
In other words, we can realized nonreciprocal 1PB with high transmission coefficient in different directions with suitable phase $\theta$: that is, $a\to c$ for $\theta=3\pi/4$, and $c\to a$ for $\theta=\pi/4$.

We noted that the nonreciprocity based on quantum interference have also been proposed via the synthetic magnetism in a cyclic three-mode system~\cite{Xu15Optical,Xu20Nonreciprocity,Chen21Synthetic}.
The condition of interference is modulated by the phase differences of the driving lasers, and the nonreciprocity is almost not affected by the loss of the auxiliary cavity.
Meanwhile, we also note that by operating dissipative couplings, nonreciprocal magnon blockade has been demonstrated~\cite{Wang22Dissipation}.
The mechanism for the nonreciprocity is inhomogeneous broadenings of the energy levels in different driving directions~\cite{Wang22Dissipation}, and therefore is different from the one we have considered.

\section{CONCLUSIONS AND OUTLOOK}\label{sec4}
In summary, we have demonstrated that the introduction of the loss of an auxiliary resonator yields an exceptional capability to achieve a novel form of nonreciprocal quantum correlations, termed nonreciprocal PB, within a dissipative three-cavity configuration comprising two nonlinear cavities and a linear cavity.
This nonreciprocity manifests in both the classical regime (nonreciprocal transmission) and the quantum regime (nonreciprocal PB).
Through the interplay of loss-induced quantum interference and the anharmonicity of the energy spectrum arising from nonlinearity, we have identified that 1PB characterized by a high transmission coefficient, and 2PB, can be selectively realized by driving the system from one direction while being inhibited from the other.
Furthermore, our findings illustrate the ability to control the direction of nonreciprocity in both classical and quantum regimes by tuning the relative phase between the two nonlinear cavities.

This work expands the exploration of loss-induced effects into the quantum nonreciprocal regime, holding potential applications in chiral and topological quantum optics.
In a broader context, our work serves as inspiration for further explorations in quantum nonreciprocity, encompassing areas such as nonreciprocal macroscopic quantum superposition~\cite{Li23Optomechanical,Li23Nonreciprocal}, quantum squeezing~\cite{Guo23Nonreciprocal}, and other types of quantum correlations~\cite{Lu21Nonreciprocity}, all achieved through loss engineering.

{\bf Acknowledgements.}
The authors thank Yong-Chun Liu, Xun-Wei Xu and Jian Tang for helpful discussions. C.L. is supported by the National Key Research and Development Program of China (Grant No.2022YFA1404104) and the National Natural Science Foundation of China (NSFC, Grant No. 12025509).
H.J. is supported by the NSFC (Grant Nos. 11935006 and 11774086), the Hunan Provincial Major SciTech Program (Grant No. 2023ZJ1010) and the Science and Technology Innovation Program of Hunan Province (Grant No. 2020RC4047).
L.-M.K. is supported by the NSFC (Grant Nos. 12247105, 12175060, and 11935006), and XJ-Lab key project (Grant No. 23XJ02001).
B.L. is supported by the NSFC (Grant No. 12347136) and the National Funded Postdoctoral Researcher Program (Grant No. GZC20231726).

\bibliographystyle{unsrt}

\end{document}